\documentclass[12pt]{article}
\usepackage{epsfig,epsf}
\renewcommand{\thefootnote}{\fnsymbol{footnote}}
\begin{document}
\def\beq{\begin{equation}}
\def\eeq{\end{equation}}
\def\eq#1{{Eq.~(\ref{#1})}}
\def\fig#1{{Fig.~\ref{#1}}}
\newcommand{\as}{\alpha_S}
\newcommand{\bra}[1]{\langle #1 |}
\newcommand{\ket}[1]{|#1\rangle}
\newcommand{\bracket}[2]{\langle #1|#2\rangle}
\newcommand{\intp}[1]{\int \frac{d^4 #1}{(2\pi)^4}}
\newcommand{\mn}{{\mu\nu}}
\newcommand{\tr}{{\rm tr}}
\newcommand{\Tr}{{\rm Tr}}
\newcommand{\T} {\mbox{T}}
\newcommand{\braket}[2]{\langle #1|#2\rangle}
\newcommand{\ab}{\bar{\alpha}_S}

\setcounter{secnumdepth}{7}
\setcounter{tocdepth}{7}
\parskip=\itemsep               %?

\setlength{\itemsep}{0pt}       %?
\setlength{\partopsep}{0pt}     %?
\setlength{\topsep}{0pt}        %?
%---layout fuer eine dina4 seite-------------------
\setlength{\textheight}{22cm}
\setlength{\textwidth}{174mm}
\setlength{\topmargin}{-1.5cm}

%\input psfig
%%%%%%%%%%%%%%%%%%%%%%%%%%%%%%%%%%%%%%%%%%%%%%%%%%%%%%%%%%%%%%%
%
\renewcommand{\thefootnote}{\fnsymbol{footnote}}
\newcommand{\beqar}[1]{\begin{eqnarray}\label{#1}}
\newcommand{\eeqar}{\end{eqnarray}}
\newcommand{\m}{\marginpar{*}}
\newcommand{\lash}[1]{\not\! #1 \,}
\newcommand{\nn}{\nonumber}
\newcommand{\D}{\partial}
\newcommand{\g}{{\rm g}}
\newcommand{\el}{{\cal L}}
\newcommand{\A}{{\cal A}}
\newcommand{\Ka}{{\cal K}}
\newcommand{\al}{\alpha}
\newcommand{\be}{\beta}
\newcommand{\ep}{\varepsilon}
\newcommand{\ga}{\gamma}
\newcommand{\de}{\delta}
\newcommand{\De}{\Delta}
\newcommand{\et}{\eta}
\newcommand{\ka}{\vec{\kappa}}
\newcommand{\la}{\lambda}
\newcommand{\ph}{\varphi}
\newcommand{\si}{\sigma}
\newcommand{\ro}{\varrho}
\newcommand{\Ga}{\Gamma} 
\newcommand{\om}{\omega}
\newcommand{\La}{\Lambda}  
\newcommand{\tG}{\tilde{G}}
\renewcommand{\theequation}{\thesection.\arabic{equation}}

%%%%%%%%%%%%%%%%%%%%%%%%%%%%%%%%%%%%%%%%%%%%%%%%%%%%%%%%%%%%%%%%%
% ABBREVIATED JOURNAL NAMES  
%
\def\ap#1#2#3{     {\it Ann. Phys. (NY) }{\bf #1} (19#2) #3}
\def\arnps#1#2#3{  {\it Ann. Rev. Nucl. Part. Sci. }{\bf #1} (19#2) #3}
\def\npb#1#2#3{    {\it Nucl. Phys. }{\bf B#1} (19#2) #3}
\def\plb#1#2#3{    {\it Phys. Lett. }{\bf B#1} (19#2) #3}
\def\prd#1#2#3{    {\it Phys. Rev. }{\bf D#1} (19#2) #3}
\def\prep#1#2#3{   {\it Phys. Rep. }{\bf #1} (19#2) #3}
\def\prl#1#2#3{    {\it Phys. Rev. Lett. }{\bf #1} (19#2) #3}
\def\ptp#1#2#3{    {\it Prog. Theor. Phys. }{\bf #1} (19#2) #3}
\def\rmp#1#2#3{    {\it Rev. Mod. Phys. }{\bf #1} (19#2) #3}
\def\zpc#1#2#3{    {\it Z. Phys. }{\bf C#1} (19#2) #3}
\def\mpla#1#2#3{   {\it Mod. Phys. Lett. }{\bf A#1} (19#2) #3}
\def\nc#1#2#3{     {\it Nuovo Cim. }{\bf #1} (19#2) #3}
\def\yf#1#2#3{     {\it Yad. Fiz. }{\bf #1} (19#2) #3}
\def\sjnp#1#2#3{   {\it Sov. J. Nucl. Phys. }{\bf #1} (19#2) #3}
\def\jetp#1#2#3{   {\it Sov. Phys. }{JETP }{\bf #1} (19#2) #3}
\def\jetpl#1#2#3{  {\it JETP Lett. }{\bf #1} (19#2) #3}
%%%%%%%%% notice the parenthesys is only on one side
\def\ppsjnp#1#2#3{ {\it (Sov. J. Nucl. Phys. }{\bf #1} (19#2) #3}
\def\ppjetp#1#2#3{ {\it (Sov. Phys. JETP }{\bf #1} (19#2) #3}
\def\ppjetpl#1#2#3{{\it (JETP Lett. }{\bf #1} (19#2) #3} 
\def\zetf#1#2#3{   {\it Zh. ETF }{\bf #1}(19#2) #3}
\def\cmp#1#2#3{    {\it Comm. Math. Phys. }{\bf #1} (19#2) #3}
\def\cpc#1#2#3{    {\it Comp. Phys. Commun. }{\bf #1} (19#2) #3}
\def\dis#1#2{      {\it Dissertation, }{\sf #1 } 19#2}
\def\dip#1#2#3{    {\it Diplomarbeit, }{\sf #1 #2} 19#3 }
\def\ib#1#2#3{     {\it ibid. }{\bf #1} (19#2) #3}
\def\jpg#1#2#3{        {\it J. Phys}. {\bf G#1}#2#3}  
\newcommand{\bas}{\bar{\alpha}_S}
%%%%%%%%%%%%%%%%%%%%%%%%%%%%%%%%%%%%%%%%%%%%%%%%%%%%%%%%%%%%%%%%%%%%%
%
%\renewcommand{\thefigure}{{\protect\bf\arabic{figure}}}
%
%
%
%
%\begin{titlepage}
\noindent
\begin{flushright}
\parbox[t]{10em}{
DESY-02-206\\
 TAUP- 2727-2002\\
{\bf \today} }\\
{\bf hep-ph/0211348}
\end{flushright}
\vspace{1cm}
\begin{center}
{{\Huge  \bf  QCD Saturation and $\gamma^{*}$-$\gamma^{*}$
Scattering}
\vskip1cm
{\large \bf  Michael  ~Kozlov ${}^{a)}$ \footnote{Email:
kozlov@post.tau.ac.il
.} and Eugene ~Levin ${}^{a),b)}$\footnote{Email:
leving@post.tau.ac.il, levin@mail.desy.de.}}}
\vskip1cm

{\it ${}^{ a}$)\,\, HEP Department}\\
{\it School of Physics and Astronomy}\\
{\it Raymond and Beverly Sackler Faculty of Exact Science}\\
{\it Tel Aviv University, Tel Aviv, 69978, Israel}\\
\vskip0.3cm
{\it ${}^{b)}$ DESY Theory Group}\\
{\it 22603, Hamburg, Germany}

\end{center}  
\bigskip
\begin{abstract} 	
Two photon collisions at high energy have  an important  theoretical
advantage: the simplicity of the initial state, which gives us a unique
opportunity to calculate these processes for large virtualities of both
photons in 
perturbative QCD approach. In this paper we study  QCD saturation in
two photon collisions in the framework of the Glauber-Mueller approach.
The
Glauber-Mueller formula is derived emphasizing the impact parameter
dependence ($b_t$)  of the dipole-dipole amplitude. It is shown that
non-perturbative QCD contributions are needed to describe large
$b_t $-behaviour, and the way how to deal with them is suggested.
Our approach can be viewed as the model for the saturation in which
the entire impact
parameter dependence is  determined by the initial conditions.
 The
unitarity bound for the total cross section, its energy dependence as well
as
predictions for future experiments are discussed. 

It is argued that the total cross section increases faster than any power
of $ \ln(1/x)$  in a
wide range of energy or $x$ , namely $\sigma(\gamma^*-\gamma^*)
\,\propto\,(1/Q^2) \,\,exp( a \sqrt{\ln(1/x)})\,\leq\,1/m^2_{\pi}$ where
$exp( a \sqrt{\ln(1/x)})$ reflects the $x$ dependence of the gluon density
$xG\,\propto \,exp( 2 \,a \sqrt{\ln(1/x)})$ and $m_{\pi}$ is the pion
mass.
\end{abstract}

%*********************************************************************************  
%*********************************************************************************  
\newpage 

\section{Introduction.}
\label{sec:Introduction}

	Two photon collisions at high energy have three theoretical
advantages
 over hadronic collisions or/and deep inelastic scattering: 
\begin{itemize}
	\item The simplicity of the initial state, 
which allows processes, such as large transverse  
 momentum hadronic jet production, to be calculated exactly to lowest
order in
 perturbative theory. With the advent of high quality experimental data,
 theoretical  analysis also focus on higher order corrections to the basic
 processes which  can provide an  interesting test of the
theory\cite{2PH};  
	\item  Scattering  of two photons with large but
 equal virtualities  gives  unique access to BFKL emission \cite{BFKL}, 
making this
process
 very useful for studying the dynamics
\cite{BRL,BHS,BOO,KM,KMR,EF,GLMN}; 
	\item Scattering of two virtual photons with large
  virtualities allows one to study  shadowing (screening) 
corrections on the solid theoretical basis of perturbative QCD \cite{TKM}.
\end{itemize} 

\begin{figure}[htbp]
\begin{minipage}{10.0cm}
\epsfig{file= 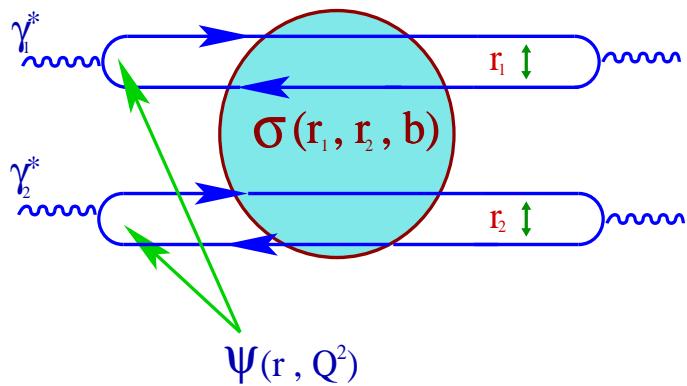,width=90mm, height=50mm}
\end{minipage}
\begin{minipage}{6.0 cm}
\caption{The picture of interaction of two  photons with virtualities
$Q_1$ and $Q_2$ larger than a  ``soft" scale.}
\label{ddint}
\end{minipage}
\end{figure}

It is well known that the correct degrees of freedom at high energy are
not quarks or gluon but  colour dipoles \cite{MU90,LR87,KOP,MU94} which
have transverse sizes $r_t$ and the fraction of energy $x$.
Therefore, two photon interactions occur in two successive   steps.
First, each of
virtual photon decays into a colour dipole ( quark - antiquark pair ) with
size $r_t$. At large value of virtualities the probability of such a
decay can be calculated in pQCD. The second stage is the interaction of
colour dipoles with each other. The simple formula ( see for example
Ref. \cite{DDR} ) that describes the
process of interaction of two photons with virtualities $Q_1$ and $Q_2$  
is  (see \fig{ddint} ) 
\beq \label{PPS}
\si(Q_1,Q_2,W)\,\,=\,\,
\int\,d^2 b_t
\sum^{N_f}_{a,b}\,\
\eeq
$$
\int^1_0 \,d\,z_1 \,\int\,d^2 d r_{1,t} |
\Psi^a_{T,L}(Q_1;z_1,r_{1, t})|^2\,\,\int^1_0 \,d\,z_2 \,\int\,d^2 d
r_{2,t} |
\Psi^b_{T,L}(Q_2; z_2,r_{2, t})|^2\,\,\si^{dd}_{a,b}
(\tilde{x}_{ab},r_{1,t},
r_{2,t}; b_t)
$$
where the indices $a$ and $b$ specify the flavours of interacting quarks, 
$T$ and $L$ indicate the polarization of the interacting photons. The
$r_i$ denote the transverse separation between quark and antiquark in the
dipole (
dipole size) and $z_i$ are the energy fractions of the quark in the photon
$i$.   $\si^{dd}_{ab}\,\,=\,\,2 N((\tilde{x}_{ab},r_{1,t},r_{2,t}; b_t)$
 where $N$ is the imaginary part of the dipole - dipole amplitude 
at
energy $x$ given by
\beq \label{X}
\tilde{x}_{ab} \,\,=\,\,\frac{Q^2_1\,+\,Q^2_2 \,+\,
4\,m^2_a\,+\,4\,m^2_b}{W^2\,+\,Q^2_1\,+\,Q^2_2}
\eeq
where $m_a$ is the mass of the quark with flavour $a$. $b_t$ is the impact
parameter for dipole-dipole interaction and it is equal the transverse
distance between the dipole centers of mass. It is clear that
$\si^{dd}_{ab}$  has a  meaning of $d \sigma/d^2 b_t$.

The wave functions for virtual photon is known \cite{WF} and they are
given by

\begin{eqnarray}
\Psi^a_{T}(Q; z,r_{ t})|^2 &=& \frac{6
\alpha_{em}}{\pi^2}\,Z^2_a\,\left(\,(z^2\,+\,(1 -
z)^2)\,\bar{Q}^2_a\,K^2_1 (\bar{Q}_a\,r_t)\,+\,m^2_a\,K^2_0
(\bar{Q}_a\,r_t)\,\right)\label{WFT}\\
\Psi^a_{L}(Q; z,r_{ t})|^2
&=&\frac{6\alpha_{em}}{\pi^2}\,Z^2_a\,Q^2\,z^2\,(1 - z)^2
K^2_0(\bar{Q}_a\,r_t)\,\label{WFL}
\end{eqnarray}
with $ \bar{Q}^2_a \,\,=\,\,z(1-z)Q^2\,+\,m^2_a$ where
$Z_a$ and $m_a$ denote the faction of charge and mass of the quark of
flavour $a$.

The main contribution in \eq{PPS} is concentrated at $r_{1,t}\,
\approx\,1/Q_1 \,\ll \,1/\mu$ and $ r_{1,t}\,\approx\,1/Q_2\,\ll\,1/\mu$ 
where $\mu$ is the soft mass scale. Therefore, at first sight, we can
safely use pQCD for calculation of the dipole-dipole cross section $\si$
in \eq{PPS}. The objective of this paper is to investigate the
dipole-dipole cross section at high energy ( low $x$ ) where  QCD
saturation is expected \cite{GLR,MUQI,MV}.  The first analysis based on
Golec-Biernat and W\"{u}sthoff model \cite{GBW} has been performed in Ref.
\cite{TKM}. Here we will extend this analysis by using the Glauber -
Mueller approach \cite{MU90,LR87,KOP} with special focus on the impact
parameter dependence which was completely omitted in the GBW model as well
as in Ref. \cite{TKM}. 

In the next section  we discuss the dipole-dipole
interaction in the Born approximation of pQCD. We show that this
approximation leads to $\si$ which decreases  as a power of $b_t$. It
turns out that $\si \,\,\rightarrow\,\,1/b^4_t$ for large $b_t
\,>\,r_{1,t}$ and $r_{2,t}$.  Of course, such a behaviour will not change
its character in higher orders of pQCD ( see Refs \cite{KW1,KW2,KW3,KW4} )
since it is a direct consequence of the massless gluon in QCD. Using the
Born approximation as the example we consider the non-perturbative
contribution that provides  an exponential decrease at large values of
$b_t \,>\,1/m_{\pi}$.

Section 3 is devoted to Glauber - Mueller formula in the case of the DGLAP
emission \cite{DGLAP}. Here, we use the advantage of  photon - photon
scattering with large photon virtualities, since we can calculate the
gluon
density without uncertainties related to non-perturbative initial
distributions in hadronic target. It is well known that no  $b_t$
dependence is induced by    DGLAP emission at least for large values 
of the impact parameter. Therefore, the entire impact parameter
dependence is due to the Born Approximation cross section. In other
words, we can use our approach as an explicit illustration of the point of
view that the non-perturbative large $b_t \,\geq\,1/2 m_{\pi}$, where
$m_{\pi}$ is the pion mass,  is determined by the initial condition (see
Refs. \cite{LRREV,FIIM}) in the contrast with the notion that such a
behaviour could change the kernel of the non-linear equation that governs
 evolution in the saturation region \cite{KW1,KW2,KW3,KW4}.

The unitarity bounds as well as different regimes of the energy behaviour
of two photon total cross sections are considered in section 4.

In section 5 we give our estimates for the values of the total cross
sections for accessible range of energies.

In the last section we summarize our results.

\section{Dipole-dipole interaction in the Born approximation.}
\label{sec:DipoleDipoleInteraction}

\begin{figure}
\begin{center}
\epsfig{file=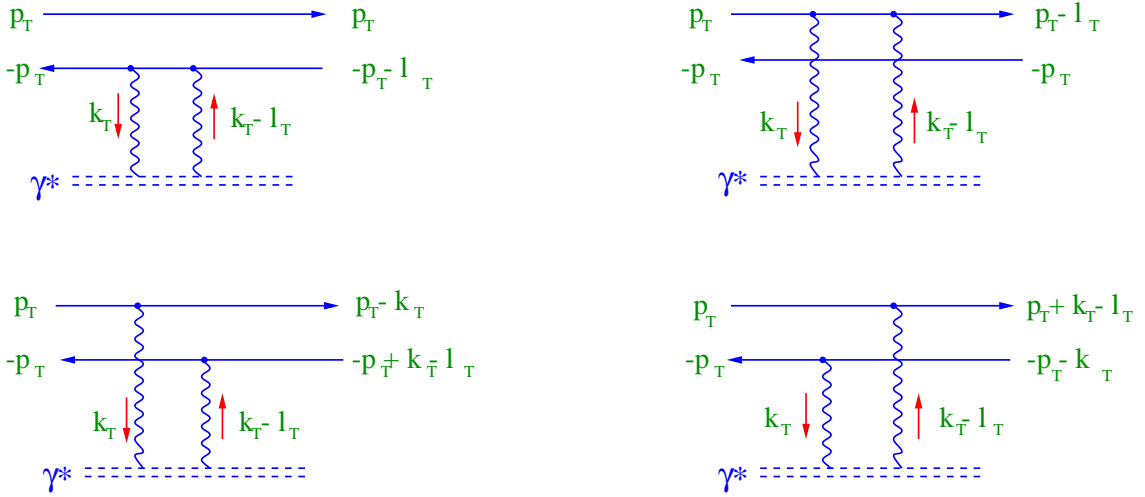,width=150mm}
\end{center}
\caption{The Born approximation for dipole-dipole scattering amplitude.}
\label{ba}
\end{figure}

The Born approximation for the dipole-dipole scattering amplitude is
shown in \fig{ba}.

To obtain the expression for $\si^{dd} (\tilde{x},r_{1, t},r_{2,t t};
b_t)$ (see \fig{ddint} ) we need to calculate the diagrams in the  
momentum
representation and than  to rewrite them in space-time
representations. The conjugated variables to $p_t$ and $l_t$ will be the
size of the dipole (say $r_{1, t}$ and the impact parameter $b_t$. The
detailed calculation performed in light-cone technique ( see for example
Ref. \cite{BRLE} )  has been performed in \cite{BAA}.  The answer is  
\beq \label{DDXS}
\si(\tilde{x}, r_{1, t},r_{2, t}; b_t) \,\,=\,\, \,\pi \as^2
\frac{N^2_c - 1}{2\,N^2_c}\,\,\left( \,\ln
\frac{(\vec{b}\,-\,z_1 \vec{r}_1 \,-\, z_2\vec{r}_2  )^2\,\,
(\vec{b}\,-\,\bar{z}_1 \vec{r}_1\, -\, \bar{z}_2 \vec{r}_2 )^2}
{(\vec{b}\,- \,\bar{z}_1 \vec{r}_1 \,- \,z_2  \vec{r}_2 )^2\,\,
(\vec{b}\,- \,z_1 \vec{r}_1\, - \,\bar{z}_2  \vec{r}_2)^2}\,\right)^2
\eeq
where $z_i$ is the fraction of the energy of the dipole carried
by quarks and $\bar{z}_i \,=\,z_i\,-\,1$. All vectors
are two dimensional in \eq{DDXS}.

\eq{DDXS} has a simpler form if we assume that $z_i = 1/2$. Namely,
\beq \label{DDXSS}
\si(\tilde{x}, r_{1, t},r_{2, t}; b_t) \,\,=\,\, \,\pi \as^2
\frac{N^2_c - 1}{2\,N^2_c}\,\,\left(\,\ln \frac{(\vec{b}\, +\,
\vec{R})^2\,\,(\vec{b}\, - \,\vec{R})^2}{(\vec{b}\, +\,
\vec{\Sigma})^2\,\,(\vec{b}\, - \,\vec{\Sigma})^2}\,\right)
\eeq
where $\vec{R}\,=\,\frac{\vec{r}_{1,t}\,-\,\vec{r}_{2,t}}{2}$ and 
$\vec{\Sigma}\,=\,\frac{\vec{r}_{1,t}\,+\,\vec{r}_{2,t}}{2}$.
We note  that we do not find  the dipole-dipole
cross section in impact parameter representation in Ref.\cite{BAA}, but
the calculation is
so simple that we just present the answer.

To simplify our further calculations we restrict ourselves by DGLAP
emission  assuming  that $r_{1, t} $ is much smaller than $r_{2,t}$. 
It is instructive to find two different limits in \eq{DDXS}.

\begin{itemize}
\item\quad $b_t \,\gg\,r_{2, t} \,\gg\,r_{1, t}$.
Expanding \eq{DDXS} one can obtain after integration over azimuthal angle
\beq \label{LBEX}
\si(\tilde{x}, r_{1, t},r_{2, t}; b_t)\,\,\rightarrow\,\,\pi \as^2
\frac{N^2_c - 1}{ \,N^2_c}\,\,\frac{r^2_{1, t}\,\,r^2_{2, t}}{b^4_t}\,\,.
\eeq
\item\quad $b_t \,\ll\,r_{1, t} \,\ll\,r_{2, t}$.
\beq \label{SBEX}
\si(\tilde{x}, r_{1, t},r_{2, t}; b_t)\,\,\rightarrow\,\,\pi \as^2
 \frac{N^2_c - 1}{ N^2_c}\,\,\frac{r^2_{1, t}\,\,r^2_{2, t}}{z^2_2 
\bar{z}^2_2\,r^4_{2,
t}}\,\,.
\eeq
\item\quad Therefore, we can suggest a simple formula which covers two
these limits, namely
\beq \label{BAPR}
\si(\tilde{x}, r_{1, t},r_{2, t}; b_t)\,\,=\,\,\pi \as^2
 \frac{N^2_c - 1}{ N^2_c} \frac{r^2_{1, t}\,\,r^2_{2,
t}}{(\,b^2_t\,+\,z_2 \,\bar{z}_2 \,r^2_{2, t}\,)^2}\,\,.
\eeq
For further estimates we will use \eq{BAPR} which reflects all
qualitative features of the full expression of \eq{DDXS} but considerably  
simplifies the calculations.

\end{itemize}

\eq{DDXS} as well as \eq{BAPR} leads to power - like decrease at large
values of $b_t$, namely, $\si(\tilde{x}, r_{1, t},r_{2, t};
b_t)\,\,\propto\,\,\frac{1}{b^4_t}$. Such behaviour cannot be
correct
since it contradicts to the general consequence of analyticity and
crossing symmetry of the scattering amplitude. Since the spectrum of
hadrons  has no particles with mass zero,  the scattering amplitude should
decrease  as $e^{ - 2m_{\pi}\,b_t}$ \cite{FROI}. Certainly we need to take
into account non-perturbative corrections to heal this problem as has
been
noticed in many papers \cite{LRREV,KW1,KW2,KW3,KW4,FIIM}.
We suggest the procedure how to introduce such corrections which is based
on the hadron-parton duality in the spirit of the QCD sum rules
\cite{QCDSR}. This procedure consists of two steps:
(i) first, we rewrite \eq{BAPR} in the  momentum transfer representation
($t = - q^2_t$) in the form of a  dispersion relation with respect to the
mass of two gluons in $t$-channel; (ii) secondly, we claim that this
dispersion integral gives correct contribution of all hadronic states on
average. Therefore, the model for the non-perturbative contribution is the
integral over a two gluon state in the  $t$-channel but with the
restriction that
two gluon mass should be larger than the minimum mass in hadronic
states, namely, larger than  $2 m_{\pi}$. As in QCD sum rules \cite{QCDSR}
we assume that the integrand at large mass of two gluon state can be found
in perturbative QCD, while for small mass we have to include the realistic
(experimental ) spectrum of hadrons. The integration from $2 m_{\pi}$
means that we believe that we can approximate the dispersion integral even
in region of small masses by the  perturbative QCD contribution. This
procedure
corresponds to the approximation that has been used in Ref. \cite{FIIM}.
 We can also
evaluate this integral differently: to take into account the first
resonance ( glueball ) explicitly and to use pQCD approach to estimate the 
dispersion integral for masses larger than $s_0$, the  value for $s_0$
 can be taken from a  QCD sum rules calculation of the glueball spectrum
\cite{QCDSR}.  Such an  approach is closely related to one developed
in Ref. \cite{LRREV} and it appears reasonable in pure gluodynamics where
we do not have any  pions.

Rewriting \eq{BAPR} in the form
\beq \label{SIMPLE}
\si(\tilde{x}, r_{1, t},r_{2, t}; b_t)\,\,=\,\,
\frac{C(r_{1,t},r_{2,t})}{(\,b^2_t\,+\,a^2\,)^2}\,\,,
\eeq
with obvious notation, we can see that
\beq \label{QXS}
\si(\tilde{x}, r_{1, t},r_{2,
t};q^2)\,\,=\,\,C(r_{1,t},r_{2,t})\,\,\int\,\,b_t d b_t
\frac{J_0(b\,q)}{(\,b^2_t
\,+\,a^2\,)^2}\,\,=\,\,\,C(r_{1,t},r_{2,t})\,\,\frac{q}{2 a}\,K_1(a\,q)
\eeq
where $J_0$ and $K_1$ are Bessel and McDonald functions respectively.
However, we can rewrite $K_1(a\,q)$ in a different way as
\beq \label{QXS1}
q\,K_1(a\,q)\,\,=\,\,\,\int \frac{J_1(
\kappa\,a)\,\kappa^2\,d\,\kappa}{\kappa^2 \,+\,q^2}\,\,.
\eeq
The last integral (see \eq{QXS1}) gives as the dispersion relation,
namely,
\beq \label{QXS2}
\si(\tilde{x}, r_{1, t},r_{2, t}; t =
-q^2)\,\,=\,\,C(r_{1,t},r_{2,t})\,\frac{1}{2 a}
\,\,\int^{\infty}_0\,\,\frac{J_1 ( \kappa\,a)\,\kappa^2\,d \kappa}{
\kappa^2\,\,-\,\,t}
\eeq
\eq{QXS2} we replace by
\beq \label{QXS3}
\si(\tilde{x}, r_{1, t},r_{2, t}; t = 
-q^2)\,\,=\,\,C(r_{1,t},r_{2,t})\,\frac{1}{2 a}
\,\,\int^{\infty}_{2 \,m_{\pi})}\,\,\frac{J_1 ( \kappa\,a)\,\kappa^2\,d
\kappa}{
\kappa^2\,\,-\,\,t}
\eeq
accordingly to our main idea. Returning  to the impact parameter
representation we obtain
\begin{eqnarray}
\si(\tilde{x}, r_{1, t},r_{2, t};b_t)\,\,&=& 
\,\,C(r_{1,t},r_{2,t})\,\frac{1}{2
a}\,\,\int^{\infty}_{2\,m_{\pi} }\kappa^2\,d \kappa
\int^{\infty}_{0} q\,dq\,\,\frac{J_1 (
\kappa\,a)\,J_0(q\,b_t) }{
\kappa^2\,\,+\,\,q^2}\,\nonumber\\
 &=& \,\,\pi \as^2
 \frac{N^2_c - 1}{2\,  N^2_c} \,\, r^2_{1, t}\,\,r^2_{2,
t}\,\,\frac{1}{ \sqrt{z_2\,\bar{z}_2}\,\,r_{2, t}}\,\,
\int^{\infty}_{ 2\,m_{\pi}}\,\kappa^2\,d \kappa\,\,\,J_1 (
\kappa\,a)\,K_0(\kappa\,b_t) \,\,. \label{BSX}
\end{eqnarray}
One can see that $\si\,\,\propto\,\,e^{ - 2 \,m_{\pi}\,b_t}$ for $b_t
\,\,\gg\,\,1/2\,m_{\pi}$ due to the asymptotic behaviour of the McDonald
function  $K_1(\kappa\,b_t) \,\,\rightarrow\,\,e^{ - 2 \,m_{\pi}\,b_t}$ at
large $b_t$.

Therefore, the $b_t$ behaviour is: for $ 1/(2\,m_{\pi})
\,>\,b_t>
r_{1, t} $ or/and $r_{2, t}$  the dipole-dipole scattering amplitude falls
as $1/b^4_t$, but for large $b_t$  ($b_t\,>\,1/(2m_{\pi})$) we have
normal
exponential decrease as $e^{ - 2\,m_{\pi}\,b_t}$ which has a
non-perturbative origin. \eq{BSX} gives us a rather general way to take
into account the non-perturbative contribution, since in this equation we
explicitly introduce the minimum mass in the experimental hadronic
spectrum.   However, as we have mentioned above, we can expect a large 
mass for the low limit of integration in the dispersion relation of
\eq{QXS3} and \eq{BSX} ($\tilde{Q}_0 \,> \,\,m_{\pi}$) which will lead to 
$\sigma(\tilde{x}, r_{1, t},r_{2, t};b_t)$ behaviour $ \propto e^{ -
\tilde{Q}_0\,\,b_t}$.

\section{Glauber - Mueller formula.}

\begin{figure}
\begin{center}
\epsfig{file=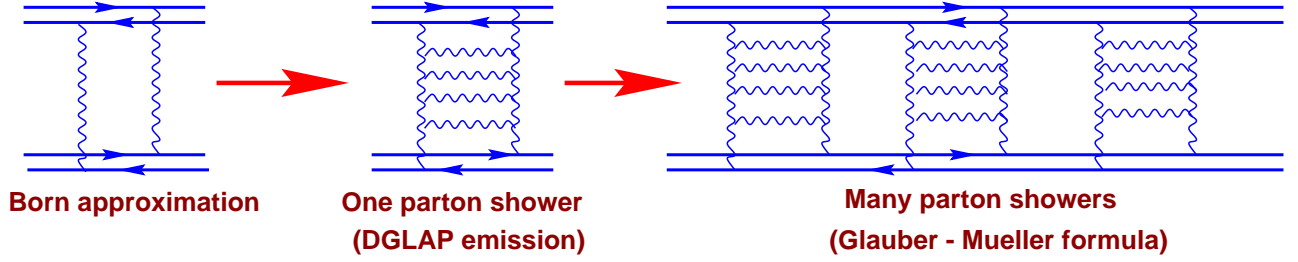,width=170mm}
\end{center}
\caption{The Glauber - Mueller  approach for dipole-dipole scattering
amplitude.} 
\label{gm}
\end{figure}

Glauber - Mueller approach takes into account the interaction of many
parton showers with the target as it is shown in \fig{gm}
Actually this formula was suggested in Refs. \cite{KOP,LR87} but Mueller
\cite{MU90} 
was the first who proved this formula especially for gluon parton density.
The main idea of this approach is that colour dipole is the correct
degree of freedom for high energy scattering \footnote{This idea was
formulated by A.H. Mueller in Ref. \cite{MU94} a bit later.}. Indeed, the
change of the value of the dipole size $r_t$ ($\Delta r_t$) during the
passage of the colour dipole through the target is proportional to
the number of rescatterings (or the size of the target $R$) multiplied by
the angle $k_t/E$ where $E$ is the energy of the dipole and $k_t$ is the
transverse momentum of the $t$-channel gluon which is emitted by the fast
dipole.

\beq \label{DOF1}
\Delta \,r_t \,\,\propto\,\,R\,\frac{k_t}{E}\,\,.
\eeq
Since $k_t$ and $r_t$ are conjugate variables and due to the uncertainty
principle 
$$
k_t\,\,\propto\,\,\frac{1}{r_t}\,\,.
$$
 
Therefore,
\beq \label{DOF2}
\Delta\,r_t\,\,\propto\,\,R\,\frac{k_t}{E}\,\,\ll\,\,r_t\,\,\,
\mbox{if}\,\,\,
\,R\,\ll\,\,r^2_t\,\,E\,\,\,\mbox{or}\,\,\,x\,\,\ll\,\,\frac{1}{2
m\,R}\,
\,.
 \eeq

\subsection{DGLAP emission}
We first  discuss  the generalization of the Born approximation to
include the DGLAP emission ( see \fig{gm} ).  This will give us the
correct 
description of the one parton shower interaction.  The DGLAP equation 
looks very simple in the region of low $x$, namely,
\beq \label{DGLAP}
\frac{\partial^2 xG(x,r^2_{1,t},r^2_{2,t})}{\partial 
\ln(1/x)\,\partial\,\ln(1/r^2_{1,t})}\,\,\,=\,\,\,\frac{N_c}{\pi}\as(r^2_{1, 
t})\,\,xG(x,r^2_{1,t},r^2_{2,t})
\,\,,
\eeq
where we consider $r_{1,t}\,\ll\,r_{2,t}$ and rewrite the DGLAP equation 
in   coordinate space. We would like to recall that the DGLAP evolution 
equation sums the $( \as \log  Q^2 )^n$ contribution, and therefore, we
can
safely rewrite it in the coordinate representation since within 
logarithmic accuracy $\ln Q^2 = \ln(1/r^2_t)$.
The initial condition for \eq{DGLAP} is $ xG(x=x_0,r^2_{1,t},r^2_{2,t})
=1$. It means 
that the dipole-dipole cross section at fixed $b_t$  for one parton shower
interaction has 
a form (
\beq \label{DIDIXS}
\sigma_{dipole}(x, r_{1,t}, r_{2,t}; 
b_t)\,\,\,=\,\,\sigma^{BA}_{dipole}(x, 
r_{1,t}, r_{2,t}; b_t) \,\,\,xG(x,r^1_{2,t},r^2_{2,t})\,\,,
\eeq          
where $\sigma^{BA}_{dipole}$ is the Born approximation for the dipole 
cross section.

The obvious solution is
\beq \label{SOLDGLAP}
xG(x,r^2_{1,t},r^2_{2,t})\,\,\,=\,\,I_0 \left( 
2\sqrt{\xi(r_{1,t},r_{2,t})\,\ln(1/x)} 
\right)
\eeq
where 
$$\xi(r_{1,t},r_{2,t})\,\,=\,\,\frac{12\,N_c}{11\,N_c\,\,-\,\,2\,N_f}\,\,
\ln\frac{\ln(4/(r^2_{1,t}\,\Lambda^2))}{\ln(4/(r^2_{2,t}\,\Lambda^2))}\,\,.
$$
 
Here,  in the arguments of the running QCD coupling we have made  the
simple 
replacement 
$ Q^2\,\rightarrow\,4/r^2_t$. Within log accuracy we cannot guarantee the 
coefficient $4$ in this expression but as was argued in Ref. \cite{GLMV}
this 
is a reasonable choice (approximation).

\eq{SOLDGLAP} has the following asymptotic behaviour
\beq \label{ASBE}
xG(x,r^2_{1,t}, r^2_{2,t})\,\,\rightarrow\,\,
e^{2\,\sqrt{\xi(r_{1,t},r_{2,t})\,\ln(1/x)}}
\eeq
which means that $ xG$ grows faster than any power of $\ln(1/x)$.

Strictly speaking \eq{DGLAP} is proven in so called double log
approximation of perturbative QCD, in which we consider
\begin{eqnarray} \label{DLA}
\as \ln (1/x) \ln (Q^2_1/Q^2_2) & \approx & 1; \nonumber\\
\as \ln (1/x) & < & 1; \nonumber\\
\as\ln (Q^2_1/Q^2_2) & < & 1; \nonumber\\
\as & \ll & 1. 
\end{eqnarray}
However, we will use this equation in a wider kinematic region where $
\as \ln (1/x) \ln (Q^2_1/Q^2_2) \,>\,1$ and $ \as \ln (1/x) \,>\,1$ while
$ \as\ln (Q^2_1/Q^2_2) \,\approx\,1$. In this kinematic region we should
use the BFKL equation \cite{BFKL}.  We view \eq{DGLAP} as the limit of the
BFKL equation in which we take into account the logarithmic contribution
in transverse momentum integration in the BFKL kernel.  The
justification  for such an approach is  the fact that the anomalous
dimension of the DGLAP equation can be parameterized in simple way
\cite{EKL}
\beq \label{ANPAR}
\gamma(\omega)\,\,=\,\,\as \,\left(\,\frac{1}{\omega}\,\,-\,\,1\,\right)
\eeq
The first term in \eq{ANPAR} leads to \eq{DGLAP} in the region of low $x$.

\eq{DIDIXS} and \eq{SOLDGLAP} solve the problem of one parton shower 
interaction in the DGLAP evolution. It should be stressed that the impact
parameter dependence enters only in the Born term in \eq{DIDIXS}. The
explanation of this fact is very simple if we recall that the logarithmic
contribution  originates from the integration over transverse momenta
large $q^2/4$ where $q^2$ is the momentum transfered along the DGLAP
ladder. Therefore, we have two choices: (i) $t\, >\, Q^2_2$ and in this
case all logs can be  summed in a function with the argument
$\ln(Q^2_1/q^2)$, (ii) $t\, <\, Q^2_2$ and in this case we have function
of
$\ln (Q^2_1/Q^2_2)$ as in \eq{SOLDGLAP}. In our problem we are certainly
dealing with the second case since we are mostly interested in large
$b_t$ - behaviour of the scattering amplitude which corresponds to low
$q^2$ behaviour.

\subsection{Many parton showers interactions}

Since the colour dipoles are correct degrees of freedom the unitarity 
constraints are for dipole-dipole  elastic amplitude 
$a_{el} (x,r_{1,t},r_{2,t};b_t)$  are diagonal and they have the form
\beq \label{UNT}
2\,\,Im a_{el} (x,r_{1,t},r_{2,t};b_t)\,\,\equiv\, 
\eeq
$$
\sigma(x,r_{1,t},r_{2,t};b_t) \,\,=\,\,|a_{el} 
(x,r_{1,t},r_{2,t};b_t)|^2\,\,+\,\,G_{in}(x,r_{1,t},r_{2,t};b_t)
$$
where $G_{in}$ stands for contribution of all inelastic processes. 
\eq{UNT} is exact for dipole-dipole scattering while it has only limited 
accuracy, for example, for dipole-proton scattering (see Ref. \cite{MU90} 
). An  experimental manifestation of the poor accuracy of \eq{UNT} for 
deep inelastic scattering is the  large cross section of so 
called inelastic diffraction dissociation of proton in an excited state.

Assuming that at high energies the amplitude is pure imaginary, one can 
find a simple solution to \eq{UNT}, namely,
\begin{eqnarray}
a_{el}(x,r_{1,t},r_{2,t};b_t)\,\,&=&\,\,i \,\left( 
1\,\,-\,\,e^{- \frac{\Omega(x,r_{1,t},r_{2,t};b_t)}{2}}\,\right)\,\,; 
\label{UNTEL}\\
G_{in}(x,r_{1,t},r_{2,t};b_t)\,\,&=&\,\,\left(
1\,\,-\,\,e^{- \Omega(x,r_{1,t},r_{2,t};b_t)}\,\right)\,\,;
\label{UNTIN}
\end{eqnarray}
where $\Omega$ is the arbitrary real function.

In Glauber - Mueller approach the opacity $\Omega$ is chosen as
\beq \label{OMEGA}
\Omega(x,r_{1,t},r_{2,t};b_t) \,\,=\,\,\sigma^{OPS}_{dipole}(x, r_{1,t}, 
r_{2,t};
b_t)\,\,\,=\,\,\sigma^{BA}_{dipole}(x,
r_{1,t}, r_{2,t}; b_t) \,\,\,xG(x,r^2_{1,t},r^1_{2,t})\,\,,
\eeq
where $\sigma^{OPS}_{dipole}$ is dipole-dipole cross section in the one
parton 
shower approximation (see \fig{gm}). One can guess that the physical 
interpretation of Glauber-Mueller formula is simple, namely, it takes 
into account  many parton shower interactions in   dipole-dipole
scattering,  
but it does not  include  a  possibility for produced partons 
from different parton showers to interact. These interactions lead to a 
more 
complicated non-linear evolution equation (see Refs. \cite{GLR,MUQI,MV,KV}
). The influence of  non-linear evolution on the photon-photon 
scattering will be discussed in a separate publication;  here we 
restrict ourselves to  consider only the first step of this non-linear 
evolution which is the Glauber-Mueller approach. 

\section{Unitarity bound}

Using the Glauber - Mueller formula of \eq{UNTEL} we can give the
unitarity
bound for dipole-dipole scattering as well as for $\gamma^*-\gamma^*$
total cross section (see
\eq{PPS}). We consider the Glauber - Mueller formula for the total
dipole-dipole cross section, namely,
\beq \label{GMXS}
\sigma^{dd}_{tot}\,\,\,=\,\,2\,\,\int
\,d^2\,b_t\,\,\left(\,1\,\,\,-\,\,\,e^{- \frac{\Omega(x,r_{1,t},
r_{2,t};b_t)}{2}}\,\right)\,,
\eeq
with opacity $\Omega $ is given in \eq{OMEGA}.

The main idea \cite{FROI} is to replace the full integration
over impact parameter in the expression for the total cross section,  by
integration in two different regions: (i) the first region is $0
\,\leq\,b_t\,\leq\,b_0(x)$; and (ii)  the second one $
b_0 \,\leq \, b_t\,\,\leq\,\,\infty$. In the first region we consider 
$\Omega /2\,\, >\,\,1$ and replace $Im\,a_{el}$ by 1. On the other hand,
in
the second region we assume that $\Omega/2 \,\,<\,\,1$ and expand
\eq{UNTEL} with respect to $\Omega$ restricting ourselves to  the first
term
of this expansion.

Therefore, 
\beq \label{UNBO}
\sigma^{dd}_{tot}\,\,\,<\,\,2\pi\,\,\left(\,\int^{b^2_0}_0\,\,d
b^2_t\,\,1\,\,\,+\,\,\int^{\infty}_{b^2_0}\,\,d\,b^2_t\,
\,\,\,\frac{\Omega}{2}\,\right)\,\,.
\eeq

\begin{boldmath}
\fbox{{$b_0\,\ll\,\frac{ 1}{2\,m_{\pi}}$}}
\end{boldmath}

Let us assume that $b_0 \,\ll\, 1/2\,m_{\pi}$ . In this case we can use
\eq{BAPR} (or \eq{SIMPLE})  for the $b_t$ - dependence for both intervals.
Taking the integral of \eq{UNBO} we have
\beq \label{UNBO1}
\sigma^{dd}_{tot}\,\,\,<\,\,2\pi \left(\,
b^2_0(x)\,\,+\,\,\frac{C(r_{1,t},r_{2,t})\,\,xG(x,r^2_{1,t},r^2_{2,t})}
{2\,(b^2_0(x)\,+\,a^2)}\,\right)\,.
\eeq
We follow Froissart's idea to evaluate the value of the characteristic  
impact parameter $b_0$, namely, the value of $b_0(x)$ can be found from
the following equation
\beq \label{VALB0}
\frac{\Omega(x, r_{1,t}, r_{2,t}; b_0(x))}{2}\,\,\,=\,\,\,1\,\,.
\eeq
Indeed, for $b_t \,>\,b_0$ $\frac{\Omega}{2} \,<\,1$  the full formula
gives less than the first term of the expansion, while for $b_t \,<\,b_0$
$\frac{\Omega}{2} \,>\,1$,  the elastic amplitude for the fixed value
of the impact parameter is less than 1. Using \eq{DIDIXS} we obtain the
solution of \eq{VALB0} in the form
\beq \label{B0}
b^2_0\,\,=\,\,-r^2_{2,t}\,(z_2\,\bar{z}_2)\,\,+\,\,\frac{
\as}{N_c}\,\sqrt{\frac{\pi\,(N^2_c -
1)}{2}}\,r_{1,t}\,r_{2,t}\,\sqrt{I_0\left(2
\,\sqrt{\xi(r_{1,t}\,r_{1,t})\,\ln(1/x)}\right)}
\eeq
 see \eq{SOLDGLAP} for all notation.

One can see that \eq{B0} leads to
$b^2_0\,\,\rightarrow\,\,e^{\sqrt{\xi(r_{1,t}\,r_{1,t})\,\ln(1/x)}}$ at
$x\,\rightarrow\,0$  which
means that $b_0(x)$ increases faster than any power of $\ln(1/x)$.

One can see that  $b^2_0$ becomes negative  at rather large values of $x$.
It reflects the fact that \eq{VALB0} does not have a solution at all
values of $x$. In other words, $\Omega/2 \,<\,1$ even at $b_t \,=\,0$ for 
low energies. However, it should be stressed that \eq{VALB0} does have a
solution at high energies which we are actually dealing with in this
paper.

Substituting \eq{B0} into \eq{UNBO} we obtain
\beq \label{UNFIN}
\sigma^{dd}_{tot}\,\,\,<\,\,2\pi
\,\left(\,2\,b^2_0(x)\,\,+\,\,r^2_{2,t}\,(z_2\,\bar{z}_2)\,\right)
\eeq
with $b_0$ of \eq{B0}. \eq{B0} is in striking contradiction with the
Froissart theorem which states that $ \sigma^{dd}_{tot} \,\,\ll
\ln^2(1/x)$ (see Ref.\cite{GLMFR} for more details on the Froissart
theorem for photon interaction.)

\begin{boldmath}
\fbox{{$b_0\,\gg\,\frac{ 1}{2\,m_{\pi}}$}}
\end{boldmath}

In this case for $b_t\,>\,b_0\,>\,\,\frac{ 1}{2\,m_{\pi}}$
the integral over $\kappa$ in \eq{BSX} is
concentrated at $ \kappa \,\rightarrow\,\,2\,m_{\pi}$ with $\kappa -
2\,m_{\pi}\,\,\approx\,\,1/b_t$. Since $2\,m_{\pi}\,a\,\ll\,\,1$ we expand
$J_1$ function, namely $J_1(\kappa\,a )\,=\,\frac{1}{2}\,\kappa\,a$. Since
we are interested only in large $b_t$ behaviour of the opacity $\Omega$
($b_0 \,\,\gg\,\,\frac{1}{2\,m_{\pi}}$) we replace $\kappa^2$ under
integral by $( 2\,m_{\pi} )^2$. The use of the asymptotic behaviour of 
McDonald's function as well as  the simplifications, mentioned above,
leads to overall accuracy $1/b_t$ in the pre-exponential factor, which is 
enough to obtain the unitarity bound. It worthwhile mentioning that in
our numerical calculation the integral of \eq{BSX} was computed without
any approximation. 

Finally we have the following estimate for the integral of
\eq{BSX}\footnote{
The integral $\int^{\infty}_{2\,m_{\pi}} z K_0(z) dz =
\,K_1(2\,m_{\pi})$. It follows directly from the differential equation for
$K_0$, namely $\frac{d}{d z} (z K_0(z)) \,\,=\,\,- z K_0(z)$ which should
be integrated over $z$. Recalling that $ - \frac{d}{d z} K_0(z) = K_1(z)$ 
we obtain the above integral. The alternative way is to use a combination
of
equations  6.561(8) and  6.561(16) from Ref. \cite{RY}. }

\begin{eqnarray}
\sigma^{BA}_{dipole}\,\,&=&\,\,\,\pi\,\as^2\,\,\frac{N^2_c -
1}{2\,N^2_c}\,(\,r^2_{1, t}\,\,r^2_{2, t}\,)
(2\,m_{\pi})^2\,\int^{\infty}_{2\,m_{\pi}}\,\kappa 
\,d\,\kappa\,\,K_0(\kappa\,b_t)\,\,;
\label{UNES1} \\
 &=&\,\,\,\pi\,\as^2\,\,\frac{N^2_c -
1}{2\,N^2_c}\,(\,r^2_{1, t}\,\,r^2_{2, t}\,)
(2\,m_{\pi})^3 \frac{1}{b_t}
\,K_1(2\,m_{\pi}\,\,b_t)\,\,;
\label{UNES2}\\
  &\rightarrow &\,\,\,\pi^2\,\as^2\,\,\frac{N^2_c -
1}{2\,N^2_c}\,(\,r^2_{1, t}\,\,r^2_{2, t}\,)
(2\,m_{\pi})^3\,\sqrt{\frac{\pi}{4\,m_{\pi}\,b^3_t}}\,e^{ -
2\,m_{\pi}\,b_t}\,\,;\label{UNES3}
\end{eqnarray}
where \eq{UNES3} gives  the asymptotic behaviour at large $b_t$ ($b_t
\,\,\gg\,\,1/(2\,m_{\pi})$ ). Namely, this is the expression we will
use for the
estimates of the value of $b_0$ in this case. Substituting \eq{UNES3} in
\eq{BSX} and \eq{OMEGA} we find the solution to  \eq{VALB0}, namely
\beq \label{B0EXP0}
b^{exp}_0(x)\,\,=\,\,\frac{1}{2\,m_{\pi}}\,\,\ln\left(
\,\pi\,\as^2\,\,\frac{N^2_c -
1}{4\,N^2_c}\,(\,r^2_{1, t}\,\,r^2_{2, t}\,)
(2\,m_{\pi})^2\,\sqrt{\frac{\pi}{4\,m_{\pi}\,b^{3, exp}_0 (x)}}
xG(x,r^2_{1,t},r^2_{2,t})\,\right)
\eeq
\eq{B0EXP0} is still an equation for $b^{exp}_0$ which has the asymptotic
solution at low $x$:
\beq \label{B0EXP}
b^{exp}_0(x)\,\,=\,\,\frac{1}{2\,m_{\pi}}\,\,\ln\left(
\,\pi\,\as^2\,\,\frac{N^2_c -
1}{4\,N^2_c}\,(\,r^2_{1, t}\,\,r^2_{2,
t}\,)\,\,(2\,m_\pi)^4 \,\sqrt{\frac{\pi}{2}}
xG(x,r^2_{1,t},r^2_{2,t})\,\right)
\eeq

One can see two important differences between this case and the case that
we have considered previously : 
\begin{enumerate}
\item \quad   $b_0$ of \eq{B0EXP} grows only logarithmically as a function
of energy. From \eq{SOLDGLAP} we conclude that 
$b^{exp}_0\,\,\propto\,\,\sqrt{\ln(1/x)}$;
\item \quad the second term in \eq{UNBO} gives a small contribution which
does not depend on energy.
\end{enumerate}

Therefore, in this kinematic region the unitarity bound has a form
\beq \label{UNBOEXPFIN}
\sigma^{dipole - dipole}_{tot}\,\,\,<\,\,2\pi\,(\,b^{exp}_0(x)\,)^2
\eeq
with $b^{exp}_0 $ from \eq{B0EXP}.

This equation reproduces the classical Froissart result \cite{FROI},
namely, the fact that the total cross section can increase only
logarithmically. 
 This is the  kind of energy behaviour we expect for DIS or hadron-hadron
collisions. However, we would like to draw your attention to the fact that
we obtain 
\beq \label{FRB}
\sigma^{dipole - dipole}_{tot}\,\,\leq\,\,
\frac{2 \pi}{(2\,m_{\pi})^2}\,\,\ln(1/x)\,\,;
\eeq
while the unitarity bound for the hadron - hadron cross section has 
$ \ln^2 s  $ - behaviour ($
\sigma^{hadron-hadron}_{tot}\,\,\leq\,\,\frac{2 \pi}{
(2\,m_{\pi})^2}\,\,\ln^2 s\,$).

It is worthwhile mentioning that \eq{FRB} holds in the wide range of the
photon virtualities which we will define below.

{\bf Predictions:}

\begin{figure}
\begin{center}
\epsfig{file=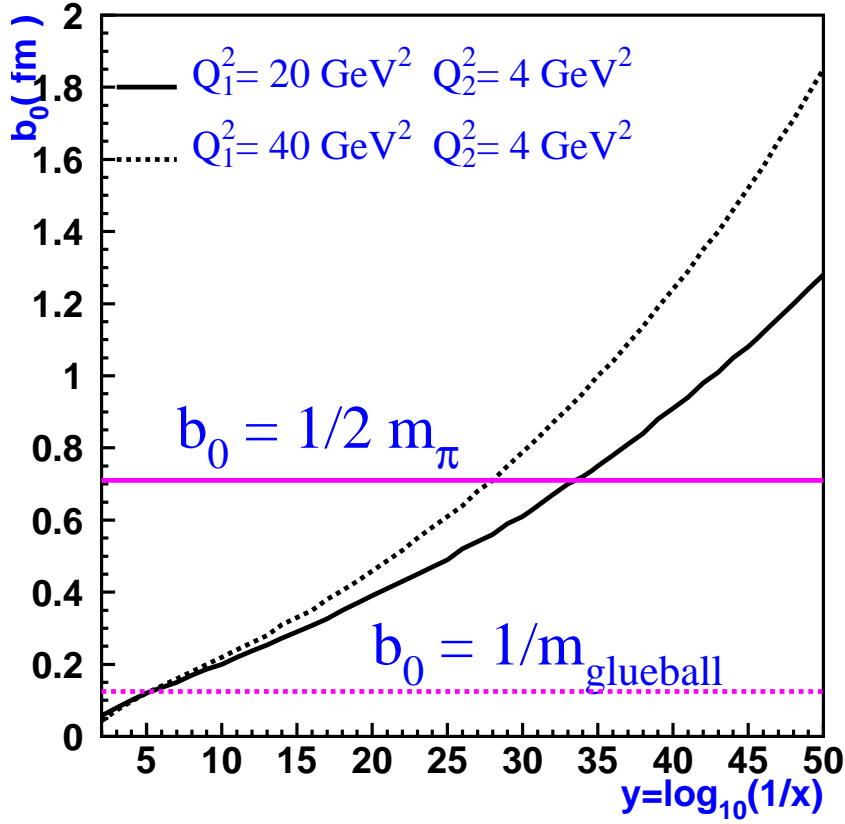,width=120mm}
\end{center}
\caption{Energy behaviour of $b_0(x)$ ( $Q^2 = 4/r^2_t$).}
\label{b0}
\end{figure}

Comparing \eq{UNFIN} and \eq{UNBOEXPFIN}, one can see that
in a  wide range of energies where $b_0(x)\,\,\leq\,\,1/2\,m_{\pi}$ the
photon - photon scattering shows an exponential ($\propto
e^{\sqrt{a \ln(1/x)}} $ ) behaviour as a
function of $\ln(1/x)$,  in striking contradiction with the DIS or/and
hadronic processes.
However, for higher energies $b^{exp}_0(x)$ reaches the value of
$1/2\,m_{\pi}$ or $1/m_{glueball}$. For higher energies the unitarity
bound becomes the one of
\eq{UNBOEXPFIN}. The numerical evaluation shown in \fig{b0} illustrate the
fact that the  kinematic region of an exponential   increase is wide,
especially if we believe that the non-perturbative corrections will
only appear  at small masses in t-channel.  Therefore, we find  that
$\gamma^* -
\gamma^*$ scattering shows quite
different behaviour than DIS and hadronic processes at all accessible
energies (see \fig{b0}). However, if the typical mass in t-channel is
rather the mass of a glueball (see Ref. \cite{GB} ) the non-perturbative
corrections will stop the exponential increase as $e^{\sqrt{a \ln(1/x)}}$ 
at $x \,\approx\,10^{-5}$.

It is interesting to notice that the value of $b_0(x)$ turns out to be
larger at larger value of $Q^2$ in the region of low $x$. The reason for
such behaviour is the fast increase of the gluon density at larger values
of $Q^2$ which prevails the suppression due to extra factor $1/Q$ in
\eq{B0}. From \fig{b0} one can see that $b_0(Q^2\, = \,20 GeV^2)
\,\,<\,\,b_0(Q^2\, =\, 40 GeV^2)$ at $x\,\leq 10^{-7}$.

\section{Total $\mathbf{\gamma}^{\mathbf{*}}-\mathbf{\gamma}^{\mathbf{*}}$
for accessible energies}

\begin{figure}
\begin{center}
\epsfig{file=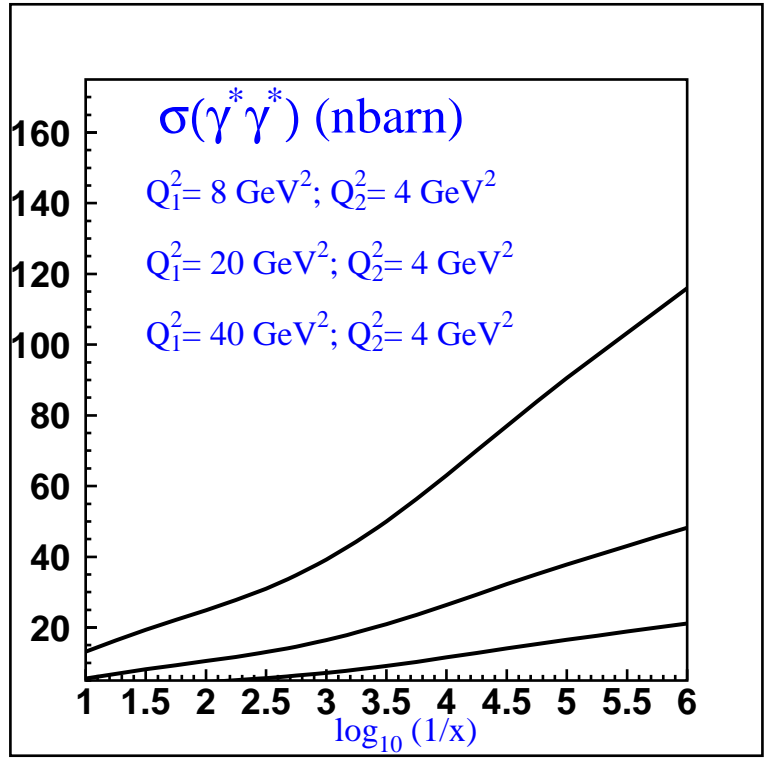,width=100mm}
\end{center}
\caption{Energy behaviour of total $\gamma^*-\gamma^*$ cross section.}
\label{re}
\end{figure}

Using the master formula of \eq{PPS} with the dipole-dipole cross section
given by \eq{GMXS} we calculate the $\gamma^*-\gamma^*$ total cross
sections at accessible range of energies. The results of calculations are
presented in \fig{re}. We fix the virtuality of  one of the  photons  at
$Q^2_2 = 4 \,GeV^2$ and calculate the cross section at different values of
$Q^2_1$.  It is essential to recall that we discuss $\gamma^*-\gamma^*$
scatterinbg in the DGLAP dynamics and we have to fix large values for
virtualities of both photons. $Q^2_2 =  4 \,GeV^2$ corresponds to
$r_{2,t}\,\,\approx\,\,0.2\,\,fm$ which is smaller than the
electromagnetic radius of pion ($R_\pi = 0.66\,\,fm$). Therefore, we can
apply perturbative QCD to our process.

\begin{figure}   
\begin{center}
\epsfig{file=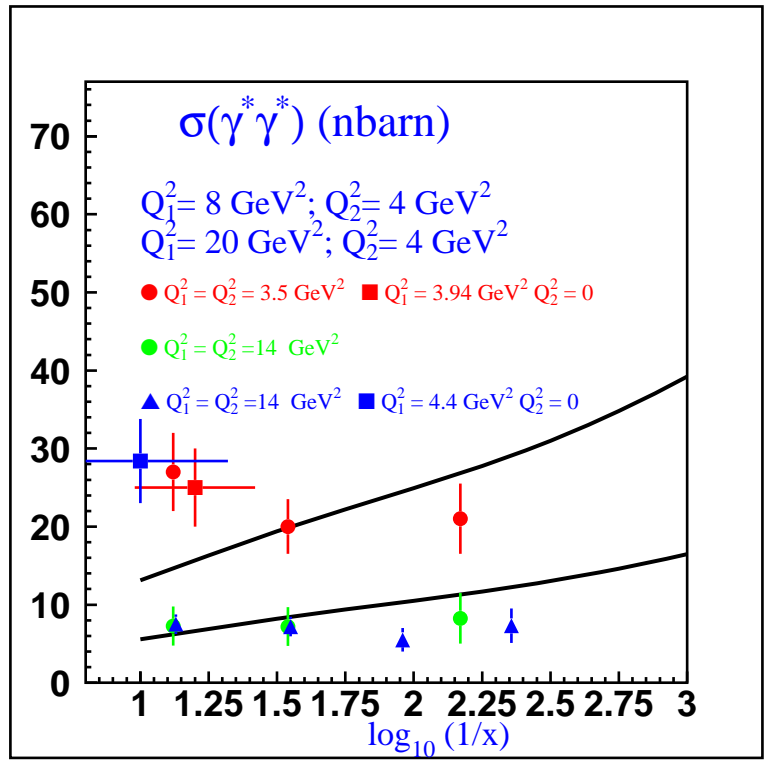,width=100mm}
\end{center}
\caption{Energy behaviour of total $\gamma^*-\gamma^*$ cross section for
low energies and experimental data.   Squares denote the L3 data
\protect\cite{L3} while the
triangles mark OPAL data \protect\cite{OPAL}. Circles label data taken from
\protect\cite{STDATA}.}
\label{re1}
\end{figure}

In \fig{re1} we also show the experimental data for the 
$\gamma^* - \gamma$ -process since there is no experimental information
about the values of the cross sections for $\gamma^* - \gamma^*$ -
scattering for large but different photon virtualities. However, the main
dependence of the cross section
is on the largest virtualities and we can hope that the data on $\gamma^*-
\gamma$ reaction is not  very different  from  $\gamma^* - \gamma^*$ one.

One can see from \fig{re1} that our predictions  are  not in contradiction
with available
but poor experimental data. 
 We see in \fig{re1} that
data with one real photon overshoot our predictions. Actually, there are
more data on $\gamma^*-
\gamma$ reaction but they are presented in the form of photon structure
function. We do not want to recalculate  the cross section using these
data since we, being theorists, are not entitled to put experimental
errors for these reconstructed data. We would like to mention once more
that this comparison with the experiment could be considered only as
illustrative one showing that we obtain a reasonable estimates for the
value of the cross sections. 
 The fact that the data with large but equal virtualities are
less than our prediction is understandable since in our approach the
 cross sections for
such processes do not have an extra enhancement due to the 
gluon structure
function.

 Therefore, we can view \fig{re1} as an argument  that our predictions do
not
contradict the current  experimental data and as the reason for  our
expectations
 that  future experiments will provide us with  data
which we will be able to compare with our predictions.

\begin{figure}[htbp]
\begin{minipage}{10.0cm}
\epsfig{file= 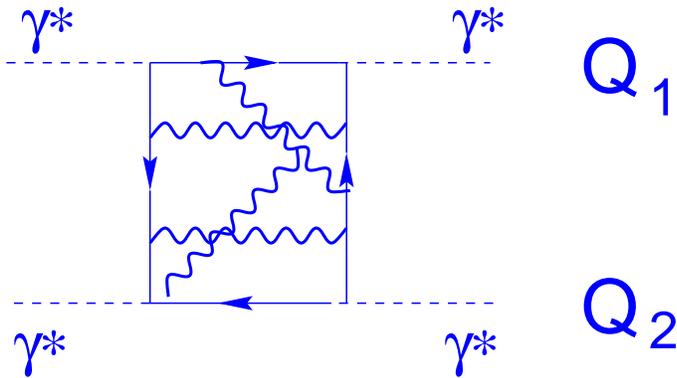,width=90mm, height=50mm}
\end{minipage}
\begin{minipage}{6.0 cm}
\caption{The picture of interaction of two  photons with virtualities
$Q_1$ and $Q_2$ due to quark-antiquark pair exchange ( so called ``box"
diagram).}
\label{box} 
\end{minipage}
\end{figure}

It should be mentioned that for serious comparison with the experimental
data we have to calculate the power-like corrections to high energy
behaviour discussed in this paper. These corrections are calculable for
$\gamma^* - \gamma^*$ scattering and they can be described as the exchange
of quark-antiquark pair in $t$-channel ( so called ``box" diagram of
\fig{box}) \cite{BOX}. The simple ``box" diagram without gluon emission
falls down as $1/W^2$ where $W$ is the energy of $\gamma^* - \gamma*$
scattering. However, the gluon emission slow down this decrease and,
therefore, such corrections could be important at sufficiently high
energies.  

\begin{figure}
\begin{center}
\epsfig{file=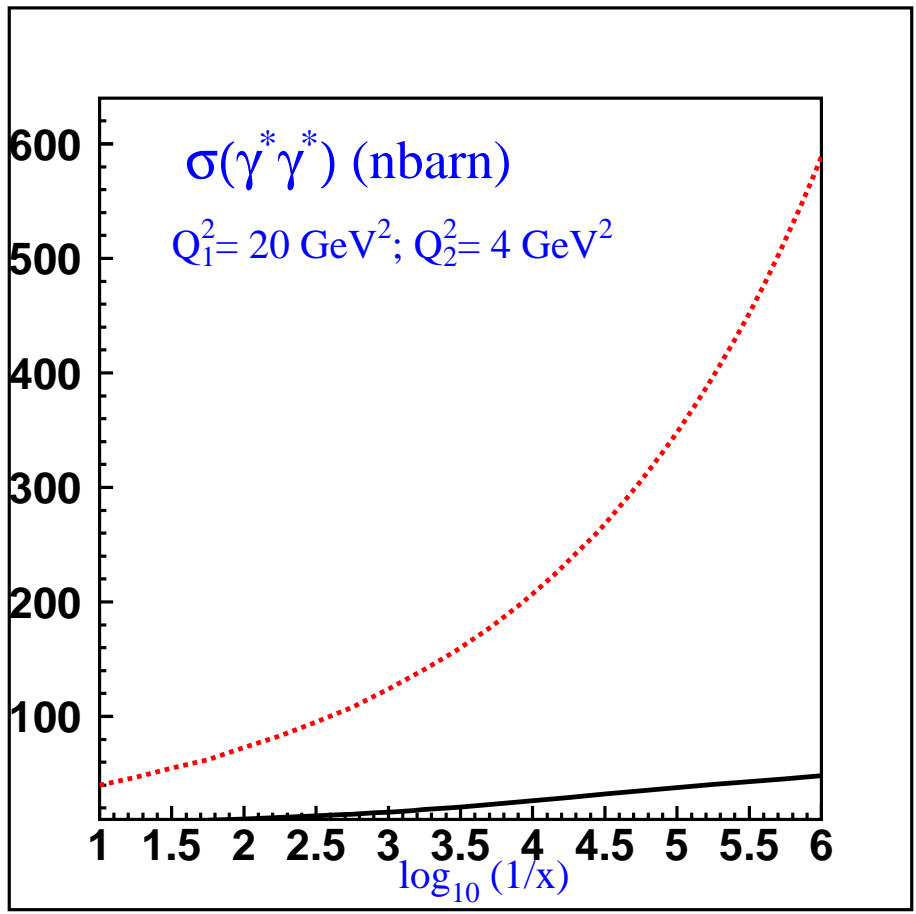,width=120mm}
\end{center}
\caption{Energy behaviour of total $\gamma^*-\gamma^*$ cross section
(solid line) and 
unitarity bound (dotted line).}
\label{unit}  
\end{figure}

It is instructive also to compare the realistic calculation 
with the unitarity bound (see
\fig{unit}). To calculate the unitarity bound we use \eq{PPS} where we
substitute
\beq \label{UNCAL}
\int\,\,d^2\,b_t \,\,\sigma^{dd}_{a,b}(x,r^2_{1,t},r^2_{2,t};b_t)\,\,=\,\,
2 \pi \,\left(\,2\,b^2_0(x)\,\,+\,\,r^2_{2,t} (z_2
\,\bar{z}_2)\,\right)\,\,.
\eeq 

 One can see that the unitarity bound considerably
overestimates the value of the
cross section.

\section{Summary}

We can summarize our approach  in the following way. The kinematic region
 which we study in this paper  is the high density QCD region. In this
region
 we have the system of partons at short distances at which $\alpha_S$ is
 small, but the density of partons has become so large that we can not
apply
 the usual methods of pQCD. The important method to deal with hdQCD is a
 Glauber- Mueller  approach, which gives the simplest approximation for
the high
parton density effects. Developing the  Glauber- Mueller  approach, we
obtained the following results.

 \begin{itemize}

\item \quad  Both  DGLAP and BFKL equations are linear evolution
equations
 predicting a steep growth of cross sections as a function of energy.
 However, it is believed that unitarity holds for all physical processes.
 At high energies it manifests
itself as a suppression of the growth of the cross section.
At the saturation scale $Q_s (x)$ nonlinear effects set in.
These  effects are due to formation of a high density parton system.
\item  \quad In this paper for the first time
 the Glauber - Mueller approach has been developed for the case of
virtual photon - photon scattering. It allows us to estimate the
saturation
scale where the transition occurs from the low density to the high parton
density
regime . The estimate is made  from the equation ( $r_{1} > r_2$)

\begin{equation} \label{satscale}
\Omega(b=0, r_{1,saturation},r_{2})/2 =1
\end{equation} 

The solution of this equation is shown in Fig. \ref{SatScale}.

\begin{figure}[h]
        \begin{center}
                \epsfig{file=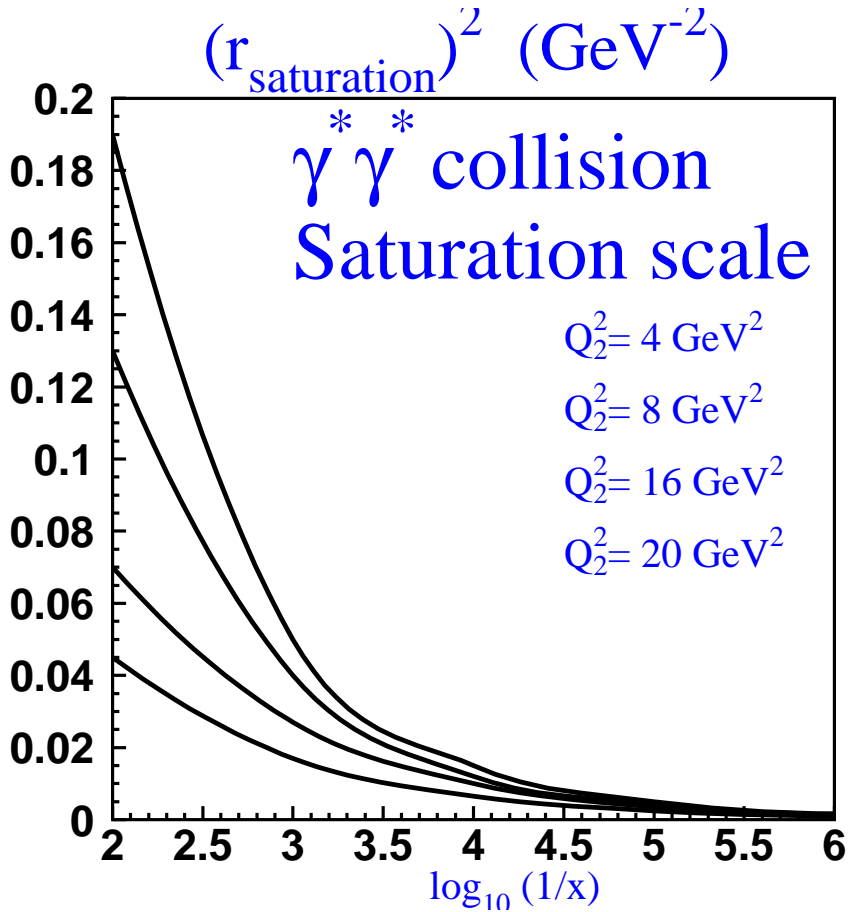, width=120mm}
        \caption{Estimate of
 the saturation radius  where the transition occurs from low density to
 high parton density regime. 
Solution given as dependence of $r^2_{saturation}$ on  $\log_{10} (1/x)$
for
fixed
values of $Q_{2}^{2}$. }
                \label{SatScale}
        \end{center}
\end{figure}

The solution to \eq{satscale} is proportional to 
$ r^2_{1,saturation}\,\,\propto\,\,r^2_{2,t} \,\,
\left(xG(x,r^2_{1, saturation},r^2_{2,t})\right)^{-1}$.

It is not surprising that the value of $r_{1,saturation}$ decreases as
function of $Q_2$ as one can see in \fig{SatScale}. At first sight it
looks strange that the value of the saturation scale $Q^2_s
\,=\,4/r^2_{saturation}$ is rather large. Indeed, for $x \,=\,10^{3}$ and
$Q^2_2 = 4\,GeV^2$ the value of  $Q^2_s \,\approx\,\,70\,GeV^2$ is much
larger than expected $Q^2_s\,\approx\,\,1 - 2\,GeV^2$ for proton.

To understand this difference we take the parameterization of the gluon
structure function for proton in the form of \eq{SOLDGLAP},namely 
$xG(x,Q^2) = G_0 I_0 (2 \,\sqrt{\xi(2/Q, R_p) \,\ln(1/x)})$ \cite{AGLPH}.
$R_p$  is a proton radius which in this estimate we can take $R^2 =
10\,GeV^{-2}$. $G_0$ is equal 0.136.  One can obtain 
$$
\frac{Q^2_s( \gamma^* - \gamma^*)}{Q^2_s( \gamma^* -
proton)}\,\,\propto\,\, Q^2_2\,R^2_p (1/G_0)\,\,\approx\,\,70\,\,.
$$

Therefore, we claim that the large value of the saturation momentum is one
of the interesting features of the $\gamma^* - \gamma^*$ scattering at
high energy.

      \item \quad We note   that the  gluon interaction leads
to power-like  decrease of the opacity ($\Omega$) in Glauber - Mueller
formula as a function of
 the impact parameter ($b$), namely $\Omega \,\propto\,1/b^4_t$.
 It turns out that because of this behavior the $\gamma^* - \gamma^*$
cross
section has a wide range of energy where it increases faster than any
power of $\ln(1/x)$ 
 in remarkable contradiction with hadron - hadron and deep inelastic cross
 sections, which cross sections can have only $\ln^2 W$  growth with
energy \cite{FROI}. This fast rise of the $\gamma^* - \gamma^*$ cross
section continue up to energies at which the typical impact parameter
($b_0(x)$) will reach the value of $1/2\,m_\pi$ ($ b_0 = 1/2\,m_\pi$ , see
\fig{b0}).

       \item \quad  The influence of this power-like $b_t$ behaviour
on
unitarity
bound
 is studied. This bound is  calculated to give an 
 estimate for the  energy behavior of the cross section.

        \item \quad It is  shown that non-perturbative contributions are
needed
 even for the case of photon-photon scattering with large virtualities of
both photons in order 
to describe the large $b_t$-behavior of the dipole-dipole scattering
amplitude.

 \item \quad  We found out that the unitarity bound for dipole-dipole  
cross section for very high energies is
$ \sigma(\gamma^* - \gamma^* )\,\,\leq\,\,\frac{2
\pi}{(2\,m_\pi)^2}\ln(1/x)$. This result can be translated in the
unitarity bound for $\gamma^* - \gamma^*$ cross section after integration
over $r_{1,t}$ and $r_{2,t}$ in \eq{PPS}. 

For $Q^2_2
\,\ll\,\,Q^2_1\,\leq\,Q^2_{1, sat} = 4/r^2_{1,saturation}$  we obtain:
\begin{eqnarray}
\sigma_{T,T}(\gamma^* - \gamma^*) & \leq & \sum_{a,b} \left(\frac{4
\alpha_{em}}{\pi})\right)^2 Z^2_a\,Z^2_b\, \ln (Q^2_{1, sat}/Q^2_1)\,\,\ln
(Q^2_{1,sat}/Q^2_2)\,\left(\frac{2\pi}{(2\,m_\pi)^2}\,\right)\,\ln(1/x)\,;
\nonumber\\
\sigma_{T,L}(\gamma^* - \gamma^*) & \leq &\sum_{a,b}
\,\left(\frac{4\alpha_{em}}{\pi})\right)\,
\left(\frac{6\alpha_{em}}{\pi})\right)\,\,Z^2_a\,Z^2_b\,\ln (Q^2_{1,
sat}/Q^2_1)\,\,\,\left(\frac{2\pi}{(2\,m_\pi)^2}\,\right)\,\ln(1/x)\,;
\nonumber\\
\sigma_{L,T}(\gamma^* - \gamma^*) &\leq & \sum_{a,b}
\,\left(\frac{4\alpha_{em}}{\pi})\right)\,
\left(\frac{6\alpha_{em}}{\pi})\right)\,\,\,Z^2_a\,Z^2_b\,\ln (Q^2_{1,
sat}/Q^2_2)\,\,\,\left(\frac{2\pi}{(2\,m_\pi)^2}\,\right)\,\ln(1/x)\,;
\nonumber\\
\sigma_{L,L}(\gamma^* - \gamma^*) & \leq & \sum_{a,b} \left(\frac{6
\alpha_{em}}{\pi})\right)^2 Z^2_a\,Z^2_b \,
\left(\frac{2\pi}{(2\,m_\pi)^2}\,\right)\,\ln(1/x)\,;
\label{PHPHUB}
\end{eqnarray}

In \eq{PHPHUB} for transverse  polarized photon we used the
logarithmic  approximation in the  integral over $r_t$. 	Indeed,
$| \Psi_T |^2\,\,\propto 1/r^2_t$ and it should be integrated from
$4/Q^2_{sat}$ to $4/Q^2$. Taking into account that $Q^2_{sat}
\,\,\propto\,\,xG\,\,\propto
\,\,e^{2\,\sqrt{\xi(r_{1,t},r_{2,t})\,\ln(1/x)}}$ ( \eq{ASBE}) one can
see  that

\begin{eqnarray}
\sigma_{T,T}(\gamma^* - \gamma^*)\,\,& \leq &\,\,\frac{C_{T,T}}{(2
\,m_\pi)^2}\,\,\ln^2(1/x)\,; \label{PPUB1}\\
\sigma_{T,L}(\gamma^* - \gamma^*)\,\,& \leq &\,\,\frac{C_{T,L}}{(2
\,m_\pi)^2}\,\,\ln^{\frac{3}{2}} (1/x)\,; \label{PPUB2}\\
\sigma_{L,L}(\gamma^* - \gamma^*)\,\,& \leq &\,\,\frac{C_{L,L}}{(2
\,m_\pi)^2}\,\,\ln^2(1/x)\,; \label{PPUB3}
\end{eqnarray}

Therefore, only $\sigma_{T,T}$ has the same energy dependence of the
 unitarity bound as hadron -hadron cross section.

 \item \quad Our approach shows that the non-perturbative corrections at
large $b_t$ should be taken into account in the Born cross section. 
Another way to treat this result is to say that the non-perturbative
corrections can be taken into account only in the initial conditions as
was discussed in Refs. \cite{LRREV,FIIM}. We do not see that such
corrections are needed in the kernel of the non-linear evolution equation
\cite{KV} as was argued in Refs. \cite{KW1,KW2,KW3,KW4}. 

  The estimates at what energies such corrections will enter the game are
  presented and discussed.

        \item  \quad Numerical calculations are performed for the value of
the
total
 cross section for accessible
energies and virtualities. These predictions will be checked soon with new 
coming data.
\end{itemize}
 
We hope that this paper will stimulate further experimental study of
$\gamma^*-\gamma^*$ - processes which can give a very conclusive
information on the saturation kinematic region in QCD.

{\bf  Acknowledgments:} We would like to thank the DESY Theory
Group for their hospitality and creative atmosphere during several
stages of this work.

We wish to thank  Jochen Bartels, Errol Gotsman and Uri Maor   for very
fruitful discussions on the subject.
                
This research was supported in part by the BSF grant $\#$
9800276, by the GIF grant $\#$ I-620-22.14/1999
  and by
Israeli Science Foundation, founded by the Israeli Academy of Science
and Humanities.

\end{document}